# Full-Range Approximation for the Theis Well Function Using Ramanujan's Series and Bounds for the Exponential Integral

Manotosh Kumbhakar[1, *] and Vijay P. Singh[1]

[1]Department of Biological and Agricultural Engineering, Texas A&M University, College Station, TX 77843, USA.

**Corresponding author details:**

Name – Manotosh Kumbhakar, Ph.D.

Email – manotosh.kumbhakar@gmail.com

[*]*Present address: Department of Civil Engineering, National Taiwan University, Taipei, Taiwan.*

**Abstract:** The solution of the governing equation representing the drawdown in a horizontal confined aquifer, where groundwater flow is unsteady, is provided in terms of the exponential integral, which is famously known as the *Well function*. For the computation of this function in practical applications, it is important to develop not only accurate but also a simple approximation that requires evaluation of the fewest possible terms. To that end, introducing Ramanujan's series expression, this work proposes a full-range approximation to the exponential integral using Ramanujan's series for the small argument ($u \leq 1$) and an approximation based on the bound of the integral for the other range ($u \in (1,100]$). The evaluation of the proposed approximation results in the most accurate formulae compared to the existing studies, which possess the maximum percentage error of 0.05%. Further, the proposed formula is much simpler to apply as it contains just the product of exponential and logarithm functions. To further check the efficiency of the proposed approximation, we consider a practical example for evaluating the discrete pumping kernel, which shows the superiority of this approximation over the others. Finally, the authors hope that the proposed efficient approximation can be useful for groundwater and hydrogeological applications.

## 1 Introduction

Pumping test analysis in groundwater flow is required for the determination of the storage coefficient and transmissivity of a confined or leaky aquifer. One can consider a horizontal confined aquifer with constant thickness, infinitely extended horizontally, and homogeneous and isotropic. Further, a single pumping well with a constant rate and a negligibly small diameter is assumed. Then, the coupling of the continuity equation and Darcy flux law gives rise to the governing partial differential equation representing the drawdown. (Walton, 1970). Theis (1935) first constructed the analytical solution for the governing equation, following an analogy with heat conduction in solids, given as follows:

$$h_0 - h(r,t) = \frac{Q}{4\pi T} \int_u^\infty \frac{\exp(-u)}{u} du = \frac{Q}{4\pi T} W(u) \qquad (1)$$

where $u = \frac{r^2 S}{4Tt}$, $r$ is the radial coordinate, $t$ is the temporal variable, $T$ and $S$ are the transmittivity and storativity, respectively; $h$ and $h_0$ are the hydraulic head and initial hydraulic head, respectively; and $Q$ is the pumping rate. The solution given by Eq. (1) is famously known as the *Theis solution* and $W(u)$ is called the *Well function*. However, in mathematics literature, $W(u)$ is known as the *exponential integral*.

Computation of the Well function is an essential part of the analysis of pump test data for confined or leaky aquifers. The direct integration of $W(u)$ is not analytically tractable. The earlier investigations used to rely on the tabular values or a few terms of the series expression (Jacob, 1940). However, these approximations are valid only for a small argument. For large values of the argument, an asymptotic (divergent) series is available, which gives an accurate approximation only within a certain range of the argument (Bleistein and Handeisman, 1975; Coulson and Duncanson, 1942; Harris, 1957). Many studies are available for finding the approximation based on polynomial or rational approximation or series expansions that are valid within a restricted domain (Allen, 1954; Cody and Thacher, 1968; Abramowitz and Stegun, 1970; Srivastava, 1995; Srivastava and Guzman-Guzman, 1998; Tseng and Lee, 1998). Swamee and Ojha (1990) combined several approximations valid for a specific region of the argument to provide an approximation to the Well function. Barry et al. (2000) constructed an approximation using the interpolation between large and small asymptotes. In a recent work, Vatankhah (2014) proposed a simple and very accurate approximation for the Well function by combining the approximations of small and large values of the argument. However, it is important to develop not only accurate but also computationally efficient approximations, which require containing the fewest possible terms of the proposed expression. In this work, we introduce S. Ramanjuan's series for the exponential integral and propose a simple and accurate closed-form formula for the Well function by combining this series and a bound.

## 2 Preliminaries and Main Results

Here, we introduce the Well function (exponential integral) and discuss its series expressions.

**Definition 1** For $u \in \mathbb{R}^+$, the exponential integral $E_1(u)$ (or $Ei(u)$) is defined as

$$E_1(u) = -Ei(-u) = \int_u^\infty \frac{\exp(-u)}{u} du \qquad (2)$$

The exponential integral $E_1(u)$ follows interesting properties, e.g., $E_1(-\infty) = -\infty$, $E_1(0) = +\infty$, $E_1(+\infty) = 0$, $E_1(u) = \Gamma(0, u)$, where $\Gamma(\blacksquare, v)$ is the upper incomplete gamma function (Abramowitz and Stegun, 1970). For evaluating the integral, one of the series expressions is given below.

**Lemma 1** (see Abramowitz and Stegun, 1970). *Let $\gamma$ denotes the Euler-Mascheroni constant. Then*

$$E_1(u) = -\gamma - \ln u - \sum_{k=1}^\infty \frac{(-1)^k u^k}{k \, k!} \qquad (3)$$

*where $\gamma=0.5772$ (up to four decimal places).*

The series expression Eq. (3) can be obtained by the power series of the exponential function and term by term integration. This series is convergent for any finite value of $u$. However, for practical application, it has some drawbacks related to the speed of convergence. Specifically, one needs to evaluate up to 75 terms of the series to obtain two significant digit accuracy over the range $u \in (0,20]$ (Coulson and Duncanson, 1942). Due to overflow and cancellation error, this series is employed only within a small range of the argument (Harris, 1957; Cody and Thacher, 1968; Stegun and Zucker, 1974; Cooper and Jacob, 1946). For the large values of $u$, the following asymptotic (divergent) series is used.

**Lemma 2** (see Bleistein and Handeisman, 1975). *For positive values of $u$, we have*

$$E_1(u) = \frac{exp(-u)}{u}\left(\sum_{k=0}^{n-1}\frac{k!}{(-u)^k} + \mathcal{O}(|u|^{-n})\right) \qquad (4)$$

*where $\mathcal{O}$ denotes the 'Big-O' notation.*

The series Eq. (4) can be obtained by expanding the integral of $E_1(u)$ by parts. This approximation can be used for practical applications only within a certain range, such as $u > 15$ or $u > 50$ (Coulson and Duncanson, 1942; Harris, 1957). Eqs. (3) and (4) are extensively used in pump test analysis for the well function approximation. Here, we introduce a faster convergent series proposed by S. Ramanujan (Berndt, 1994).

**Lemma 3** (see Berndt, 1994). *For real non-zero values of $u$, we have*

$$Ei(u) = \gamma + \ln|u| + exp(u/2)\sum_{k=1}^{\infty}\frac{(-1)^{k-1}u^k}{k!\,2^{k-1}}\sum_{n=0}^{\lfloor(k-1)/2\rfloor}\frac{1}{2n+1} \qquad (5)$$

*where $\lfloor\blacksquare\rfloor$ denotes the floor function.*

The series Eq. (5) produces faster convergence than Eq. (3), and is also valid for larger $u$ due to the term $exp(u/2)$ (Berndt, 1994). For our proposed approximation, we also take help from following inequality for the exponential integral.

**Lemma 4** *(see Gaustchi, 1959). Let us define*

$$I_q(u) = exp(u^q)\int_u^{\infty} exp(-t^q)dt \qquad (6)$$

*Then, for $q > 1$ and $u \geq 0$, we have*

$$\frac{1}{2}\left[(u^q + 2)^{1/q} - u\right] < I_q(u) \leq c_q\left[\left(u^q + \frac{1}{c_q}\right)^{1/q} - u\right] \qquad (7)$$

where $c_q = \left[\Gamma\left(1 + \frac{1}{q}\right)\right]^{q/(q-1)}$. Indeed, when $q \to \infty$, Eq. (6) becomes

$$\frac{1}{2}\ln\left(1 + \frac{2}{u}\right) \leq \exp(u)E_1(u) \leq \ln\left(1 + \frac{1}{u}\right), \qquad u \in (0, \infty) \qquad (8)$$

The proof of inequality Eq. (7) follows from the fact that $E_1(u) = \Gamma(0, u)$ and $\int_u^\infty \exp(-t^q)dt = (1/q)\Gamma(1/q, u^q)$. Now, to check the accuracy of the approximation, we use the percentage error (PE), defined as $PE\ (\%) = 100 \times \left[\left(W_{num}(u) - W_{approx}(u)\right)/W_{num}(u)\right]$, where $W_{num}(u)$ and $W_{approx}(u)$ denote the numerical and proposed approximations, respectively. For the numerical part, the MATLAB script 'integral' is utilized, which uses the global adaptive quadrature rule (Shampine, 2008). Fig. 1a shows the faster convergence of Ramanujan's series Eq. (5) over the series Eq. (3). We plot the numerical solution and the bounds given by Eq. (8) in Fig. 1b, where it is observed that $E_1(u)$ behaves like a negative exponential for large values of the argument and like a logarithmic curve for small values of the arguments.

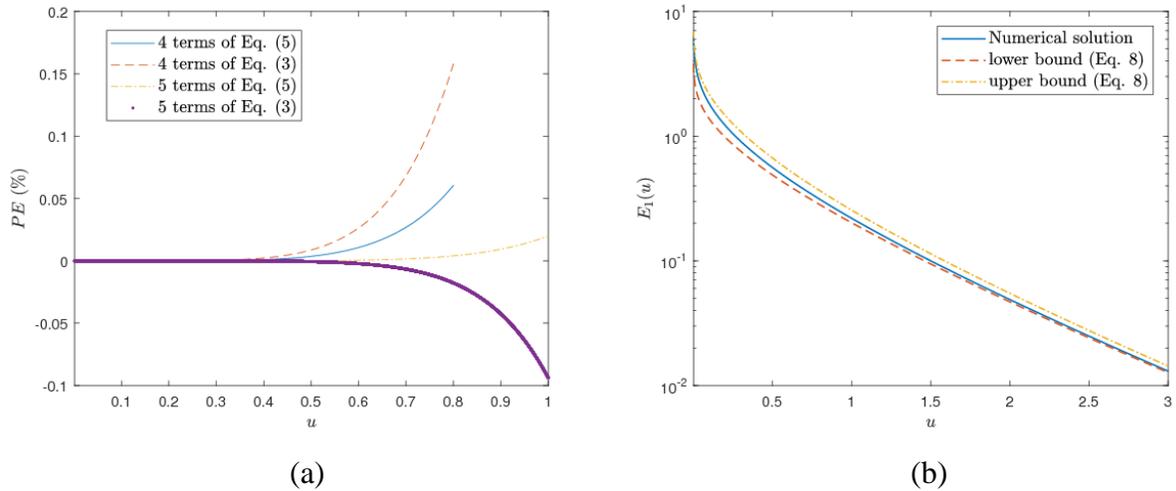

(a)          (b)

**Figure 1**: Plots of (a) Eqs. (4) and (5), and (b) Eq. (8).

Following Fig. 1, we consider series Eq. (5) with just five terms for small $u$, i.e., $u \leq 1$ and propose the following closed-form approximation for the other values of $u$, i.e., $u > 1$:

$$\ln E_1(u) = a_1 \ln\left[a_2 exp(-a_3 u) \ln\left(1 + \frac{a_4}{u^{a_5}}\right)\right] \qquad (9)$$

It may be noted that we use logarithm in Eq. (9) to deal with the small values of $E_1(u)$. Now, using the numerical solution for $E_1(u)$ and then utilizing the MATLAB curve fitting tool *cftool* along with the trust-region algorithm (Coleman and Li, 1996) to determine the coefficients $a_i$. We obtain them as $a_1 = 1.21$, $a_2 = 0.7484$, $a_3 = 0.8264$, $a_4 = 1.39$, and $a_5 = 0.8346$. Therefore, the approximation for $E_1(u)$ is obtained as follows:

$$E_1(u) = 0.7042 exp(-0.99994u)\left[\ln\left(1 + \frac{1.39}{u^{0.8346}}\right)\right]^{1.21} \quad for\ u > 1 \qquad (10)$$

As stated in Srivastava (1995), the derivative of the drawdown curve is a critical quantity for slope-matching methods. Thus, the proposed approximation of $E_1(u)$ should be able to accurately reproduce its derivative. We plot the percentage errors for $W(u)$ and its derivative in Figs. 2a and b, respectively, for the entire practical range $0 < u \leq 100$. For the purpose of comparison, we provide all the existing closed-form formulae and their errors in Table 1 (Vatankhah, 2014). It is observed that the maximum absolute PE (%) for the proposed approximation and its derivative are 0.05% and 0.06%, respectively, which is much smaller than the other formulae provided in Table 1. Further, Eq. (10) is much simpler for practical application as it contains just the product of two simpler terms compared to the existing formulae.

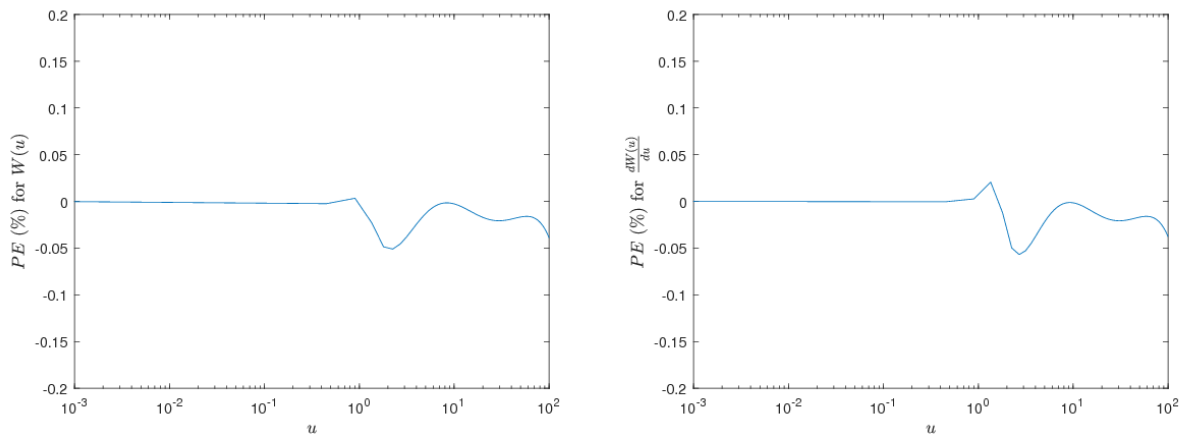

(a) (b)

**Figure 2**: Percentage errors for the proposed approximation (Eq. 5 with five terms for $u \in [0.001,1]$ and Eq. 10 for $u \in (1,100]$) of (a) $W(u)$, and (b) $dW(u)/du$.

**Table 1:** Existing approximations and PE (%) for Well function in confined aquifers over the entire range $u \in (0,100]$.

| Source | Proposed $W(u)$ | Maximum PE (%) | |
|---|---|---|---|
| | | $W(u)$ | $W'(u)$ |
| Swamee and Ojha (1990) | $\left[\left(\ln\left[(1+u)\left(\frac{0.56146}{u}+0.65\right)\right]\right)^{-7.7} + u^4 exp(7.7u)(2+u)^{3.7}\right]^{-0.13}$ | 1.28 | 2.10 |
| Barry et al. (2000) | $\dfrac{exp(-u)\ln\left[1+\dfrac{0.5615}{u} - 0.4385\left(1.0421u + \dfrac{1}{1+u^{1.5}} + \dfrac{1.0801}{1+2.35u^{-1.0919}}\right)^{-2}\right]}{0.5616 + 0.4385 exp(-2.2803u)}$ | 0.07 | 0.20 |
| Vatankhah (2014) | $\left[\dfrac{(1-0.19u^{0.7})^{-2}}{\left[\ln\left(\frac{0.565}{u}+4\right)\right]^2} + u^2 exp(2u)\left(\dfrac{u+1.384}{u+0.444}\right)^2\right]^{-0.5}$ | 0.20 | 0.22 |
| Current study | $-\gamma - \ln\|u\| + exp(-u/2)\left[u + \dfrac{u^2}{4} + \dfrac{u^3}{18} + \dfrac{u^4}{144} + \dfrac{23u^5}{28800}\right]$ for $0.001 \leq u \leq 1$<br><br>$0.7042 exp(-0.99994u)\left[\ln\left(1+\dfrac{1.39}{u^{0.8346}}\right)\right]^{1.21}$ for $u > 1$ | 0.05 | 0.06 |

## 3 Practical Application to Discrete Kernel Generator

To check the accuracy of the proposed approximation in practical applications, the discrete pumping kernel is evaluated, which is defined by (Morel-Seytoux and Daly, 1975):

$$U = \frac{1}{4\pi T}\left(W\left[\frac{r^2 S}{4tT}\right] - W\left[\frac{r^2 S}{4(t-\tau)T}\right]\right) \qquad (11)$$

where the time $t = 0$ when the well turns on and $t = \tau$ when it turns off, $\tau$ is a unit time period, and $r$ is the distance from the well. Also, Eq. (11) applies for $t > \tau$ (after the well turns off). The evaluation of Eq. (11) is important to check the proposed approximation as it involves subtraction, which might increase the errors as compared to a single approximation. Here, we consider the test case examined by previous researchers (Tseng and Lee, 1998; Barry et al., 2000; Vatankhah, 2014). Eq. (11) is evaluated for $T = 10000$ m$^2$/week, $S = 0.2$, $\tau = 1$ week, $r = 1050, 2100, 3150, 4200$ m, and $t$ varies from 2 to 18 weeks. This case corresponds to the range of $u$ as $0.30 < u < 89$. It can be observed from Fig.3 that the closed-form approximations given by Barry et al. (2000), Vatankhah (2014), and the current study work well for this realistic application over the entire practical range of the Well function. Indeed, the approximation proposed in this study is much better (maximum percentage error is about 0.1%) than the others in terms of both accuracy and simplicity.

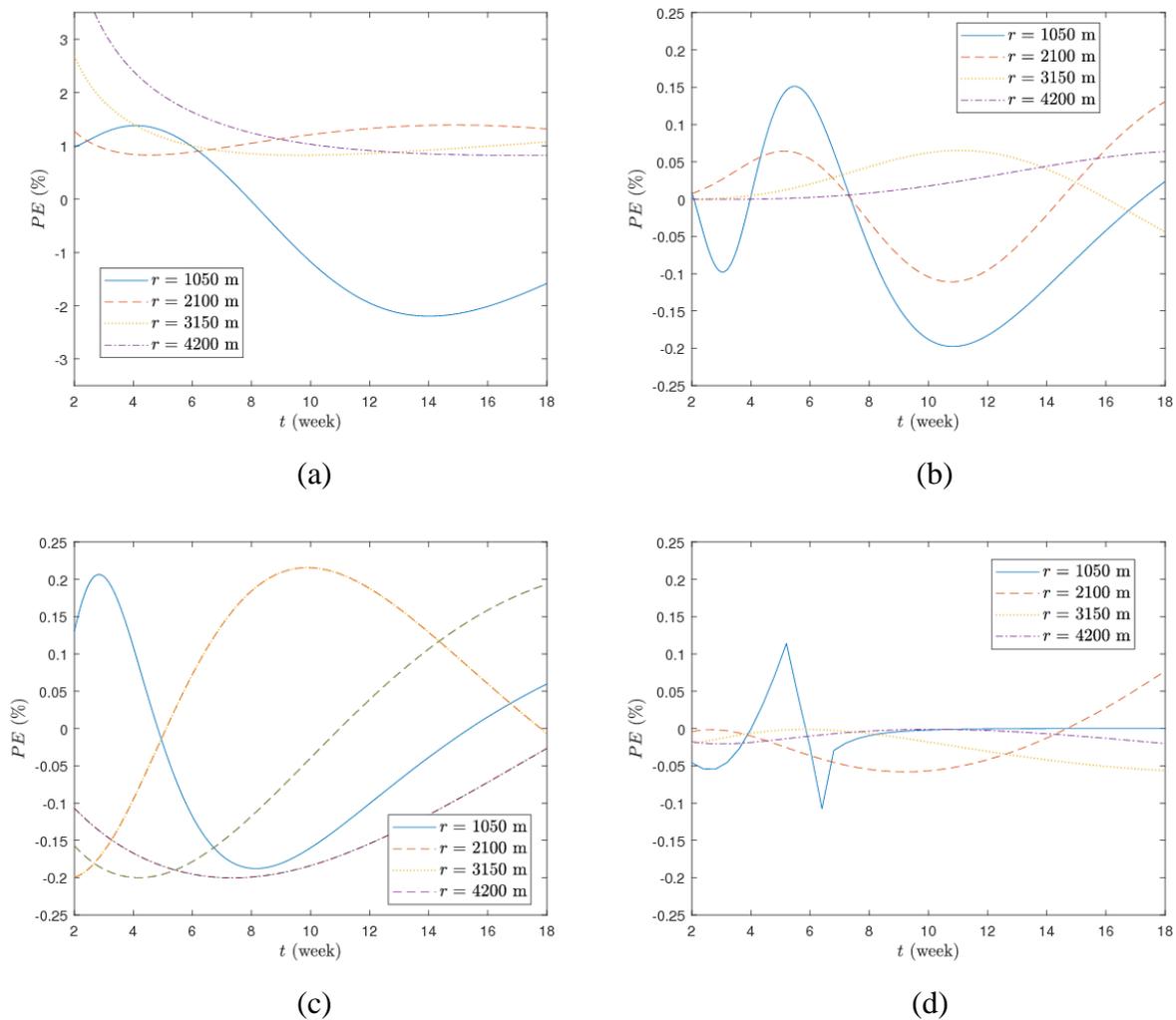

**Figure 3**: Percentage error (PE) of $U$ (week/m$^2$) as a function of time $t$ (week) for some selected from the pumping well $r$ (m): (a) Swamee and Ojha (1990), (b) Barry et al. (2000), (c) Vatankhah

(2014), and (d) Current study.

## 4 Conclusions

In the present study, we propose a closed-form formula for the Well function (exponential integral) using Ramanujan's series approximation and a bound for the exponential integral. The series approximation with just five terms is considered for the small argument, and the approximation based on the bound is seen to be well accurate for the other part of the argument. The proposed approximation is not only more accurate than the existing formulae but also is simpler in form as it requires the calculation of two terms. Further, a practical application related to the evaluation of a discrete kernel generator is considered to further check the efficiency of the proposed approximations. The results reveal the excellent accuracy of the approximation derived in this study.

**Acknowledgements** The authors dedicate this work to the mathematical genius Srinivasa Ramanujan.